# Use of Probabilistic Phrases in a Coordination Game

## Human versus GPT-4

Laurence T. Maloney[a,b*]   Maria F. Dal Martello[a,c]

Vivian Fei[a]   Valerie Ma[a]

[a]Department of Psychology, New York University
[b]Center for Neural Science, New York University
[c]Dipartmento di Psicologia Generale, Università di Padova

*Materials & Correspondence

**Laurence T. Maloney**
Department of Psychology          **Tel:** +1 (646) 512-0143
New York University               **Email:** ltm1@nyu.edu
6 Washington Place, Room 574
New York NY 10012

Acknowledgments

LTM: Partial support from the Institut d'études avancées de Paris, Paris, France. We thank Garrison Cottrell, Catherine Hanson, Stephen Hanson, and Edwin Williams for comments on earlier drafts.

Author contributions

LTM, MDM, VF and VM designed the experiment and analyses. VF and VM implemented the design as a computer-controlled experiment. LTM, MDM, VF and VM analyzed data and prepared plots and tables. LTM, MDM, VF, VM wrote the paper.

Competing interests LTM, MDM, VF and VM: none.

# Abstract

English speakers use probabilistic phrases such as *likely* to communicate information about the probability or likelihood of events. Communication is successful to the extent that the listener grasps what the speaker means to convey and, if communication is successful, individuals can potentially coordinate their actions based on shared knowledge about uncertainty. We first assessed human ability to estimate the **probability** and the **ambiguity** (imprecision) of twenty-three probabilistic phrases in a **coordination game** in two different contexts, investment advice and medical advice. We then had GPT-4 (OpenAI), a Large Language Model, complete the same tasks as the human participants. We found that GPT-4's estimates of probability both in the investment and Medical contexts were as close or closer to that of the human participants as the human participants' estimates were to one another. However, further analyses of residuals disclosed small but significant differences between human and GPT-4 performance. In particular, human probability estimates were compressed relative to those of GPT-4. Estimates of probability for both the human participants and GPT-4 were little affected by context. We propose that evaluation methods based on coordination games provide a systematic way to assess what GPT-4 and similar programs can and cannot do.

Abstract   198 words

# INTRODUCTION

English has a rich vocabulary of *probabilistic phrases* used to communicate the relative frequency or likelihood of events[1] as well as their imprecision or ambiguity (see Table 1 for examples). Your doctor might tell you that you will **probably** feel better after a night's sleep. What probability or range of probabilities do you think he intends to convey? What if instead he used the probabilistic phrase **possibly**? In the latter case the doctor is likely signalling a lower probability, but did he intentionally choose a more ambiguous term? The doctor may have reasons for introducing ambiguity. Perhaps he is not sure what effect a good night's sleep will have. Perhaps he wants to manage the patient's expectations.

**The coordination game.** In using these phrases, you, your doctor, and other English speakers you encounter are playing a simple *coordination game* (Schelling, 1960; Lewis, 2002; Franke, 2013; Benz, Ebert, Jäger & Van Rooij, 2011; Benz, Jäger & Van Rooij, 2014; Benz & Stevens, 2018) in which both players win to the extent that the probabilities and ambiguities the second player estimates are close to those the first player intended.

Human language comprises many different language games (Wittgenstein, 1953; Austin, 1955) and the successful speaker must play them all well. Our goal here is to compare human and machine performance in one of many language coordination games. Performance in the game measures how well any given player – doctor or patient – uses language as a tool (Wittgenstein, 1953; Austin, 1955; Grice, 1991). Success in playing the game would suggest that the two players in effect share an understanding of the language of probability and ambiguity that allows them to coordinate and work together well.

A coordination game is an objective task and we can compare human and machine performance trial by trial. Failure would pinpoint particular shortcomings in the performance of the machine. Can a computer play the game as well as humans do? Do successes and failures in playing the game mimic those of humans?

---

[1] We will use the term *probability* as a catch-all for relative frequency, likelihood, degree of uncertainty and probability. Following O'Brien (1989) we use the term *ambiguity* as a synonym for imprecision or ambiguity.

There is disagreement on how to assess the performance of Large Language Models such as GPT-4 (Mitchell, 2023; Mitchell & Krakauer, 2023). We follow the recommendations of Burnell et al. (2023) in our assessment: (1) We use multiple related tasks (probability and ambiguity). (2) The tasks are objective with degree of failure measured quantitatively. (3) We report exactly one run of GPT-4 – the first – in each experimental condition, not an aggregate of multiple runs. (4) We look at how GPT-4 performs in the two contexts, giving medical advice and giving investment advice.

Lastly, (5) the tasks we consider – communicating information about probability and ambiguity – are intrinsically important. There is a large literature concerning human error in decision making (Kahneman & Tversky, 1979), failures in probabilistic reasoning (Tversky & Kahneman, 1971) and the consequences of these errors and failures (Gilovich, 1993; Gilovich, Griffin & Kahneman, 2002).

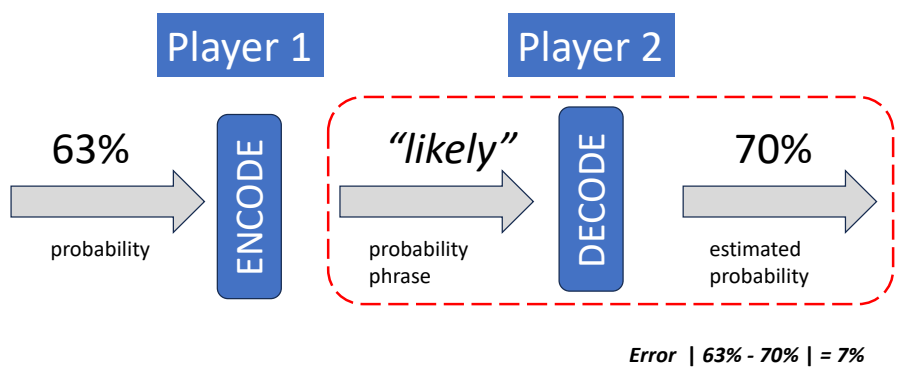

**Figure 1. The Coordination Game.** On each turn in the coordination game, the First Player is given a probability *p* (unknown to the second player) and asked to encode it as a *probabilistic phrase*. Table 1 lists the probabilistic phrases that the First Player could use to encode the probability. The Second Player (either a human participant or GPT) is then given only the selected phrase and asked to estimate the original probability. The game is a model of a transmission through a communication channel (Shannon, 1948; Shannon & Weaver, 1949). We focussed on only the Decoding phase of the game (enclosed by a dashed red contour). Either GPT-4 or a human participant played the role of Player 2 while the experimenter played the role of Player 1. In a variant of the game we asked GPT-4 or the human participant to estimate not the probability but instead the ambiguity of the probability phrase.

The capabilities of GPT-4 and other candidate Artifical General Intelligences have been compared to human with mixed results. GPT-4 fails simple intelligence tests (Biever, 2023). On the other hand, Webb, Holyoak, and Lu (2023) report that GPT-4's ability to engage in analogical reasoning and abstract pattern induction is comparable to human. Gurnee & Tegmark (2023) find that it can reason about spatial and temporal structure. GPT-4 can do more than "chat": it can write simple computer code for applications specified in natural language (Poldrack, Lu & Beguš, 2023).

| A never | B very rare | C almost never | D low risk |
|---|---|---|---|
| E low probability | F small chance | G unlikely | H there is a chance |
| I possible | J perhaps | K sometimes | L could be |
| M moderate risk | N not certain | O significant chance | P reasonable chance |
| Q reasonable to assume | R likely | S probable | T most likely |
| U expected | V almost certain | W certain | |

**Table 1: Probabilistic phrases taken from O'Brien (1989).** Each phrase has a code letter used in plotting results. Each phrase was embedded in a Medical Context and an Investment Context. Half the participants chosen at random completed the Medical Context, the remaining participants completed the Investment Context.

The probabilistic phrases we consider (Table 1) have been used in previous research with human participants (O'Brien, 1989). We modify any particular probabilistic phrase as needed so that its use in context is grammatical (e.g. *possible* can become *possibility*). In Figure 1 we illustrate one turn of the coordination game as a communication channel (Shannon, 1948; Shannon & Weaver, 1949). For simplicity we assume that, in the coordination game, the first player has only one probability and one ambiguity to signal and the second player is constrained to report a single estimate of probability and one of ambiguity.

In the full coordination game (Figure 1), Player 1 is given a target probability (for example, 63%) and must **encode** it as one of the probabilistic phrases in Table 1. Perhaps she picks *likely.* This probabilistic phrase is transmitted to Player 2 who must **decode** it and estimate the target probability. In Figure 1 she estimates 70%. The difference (7%) in absolute value between Player 1's target probability, 63%, and Player 2's estimate, %70, is the **error**, a measure of failure of coordination in the coordination game.

We will focus on the second stage (DECODE) of the coordination game (outlined in red in Figure 1), evaluating GPT-4's performance as Player 2 and comparing GPT-4's performance to that of human participants also playing as Player 2. That is, GPT-4 and its human counterpart will be asked to DECODE probability phrases and estimate corresponding probabilities. We will also ask GPT-4 and human players to rate the ambiguity (imprecision) on a scale of 0-100 of the 23 probabilistic phrases they decode. All ambiguity estimates were done after all probability estimates by both human participants and GPT-4. Probabilistic phrases were presented in randomized order to both human participants and GPT-4.

There are previous studies whose participants were asked to assign explicit probabilities to probabilistic phrases (Lichtenstein & Newman, 1967; Beyth-Marom, 1982; Bryant & Norman, 1980; Budescu & Wallsten, 1985; Kong, Barnett, Mosteller & Youtz, 1986; Mapes, 1979; O'Brien, 1989; Sawant & Sansgiry, 2018; Mellers et al, 2017; see Mosteller & Youtz, 1990 or Teigen, 2023 for review). These studies assess the extent to which humans – for the most part without any special training – agree with each other in their use of probabilistic phrases to signal probability. If all the speakers in a language community assign the same probabilities to probabilistic phrases then the players would do very well at the coordination game.

*Questions.*

*1. Can GPT-4 play the coordination game as well as humans? Are there patterned deviations between human and machine?* We needed a criterion to judge whether GPT-4, playing as Player 2, is doing what a human player would. We developed two criteria, the first based on residuals analysis (Stigler, 2016), the second on performance.

The first criterion is based on fitting a univariate linear model to data, and looking for patterned deviations (resdiduals) between human performance and that of GPT-4. These residuals may be small but highly significant. Evaluating such patterned failures is a standard part of statistical analysis (Stigler, 2016). They give us clues about shortcomings of GPT-4 even when GPT-4 overall plays the game well as measured by the second criterion. We do not claim that the linear model is an adequate model of the mapping from GPT-4's estimates of probability or ambiguity to the corresponding human estimates.

Second, we evaluate how well human and GPT-4 coordinate. We develop a measure – *discordance* – of the extent to which human players disagree with one another in playing the game and compare this measure to the discordance between GPT-4 and human players. Does GPT-4 perform as well in the coordination game as the median human player?

There is much more to language competence than assigning probability estimates and ambiguity ratings, but systematic failure to coordinate with human participants in our game would weaken any claim that GPT-4's abilities are human-like. We could not trust an Artifical General Intelligence to give medical advice if the probability phrases it uses were not correctly understood by human patients.

*2. Is GPT-4 correctly sensitive to context?* The meanings of words can depend on the context in which they occur. If your doctor and your financial consultant both use the phrase **not certain,** does it signal the same probability? Ambiguity? We will include two contexts in the experiment, medical and financial, and ask participants, including GPT-4, to rate the 23 probabilistic phrases in Table 1 for probability and for ambiguity in each context.

Keep in mind that the issue is not whether GPT-4's assessment of probability and ambiguity is invariant under context but whether GPT-4 exhibits the same changes or lack of change in probability and ambiguity ratings across context as do the human participants. In a coordination game it doesn't matter whether you are right, only whether you agree with everyone else.

*3. Is GPT-4 stable?* Lastly, we briefly investigate the *stability* of GPT-4 in this game. If we rerun the estimates by GPT-4 do we get series of similar estimates or a series of similar estimates with the occasional highly discrepant estimate? Is GPT-4 stable? We might hesitate to permit an Artificial General Intelligence to give medical advice if 1 time out of 100 it produced markedly discrepant estimates of probability or ambiguity. The motivation for testing stability will become clear when we examine the data.

# RESULTS

The results are presented in three numbered sections corresponding to the numbered questions above. We split the first question in two, one part (1A) concerned with probability, one with ambiguity (1B).

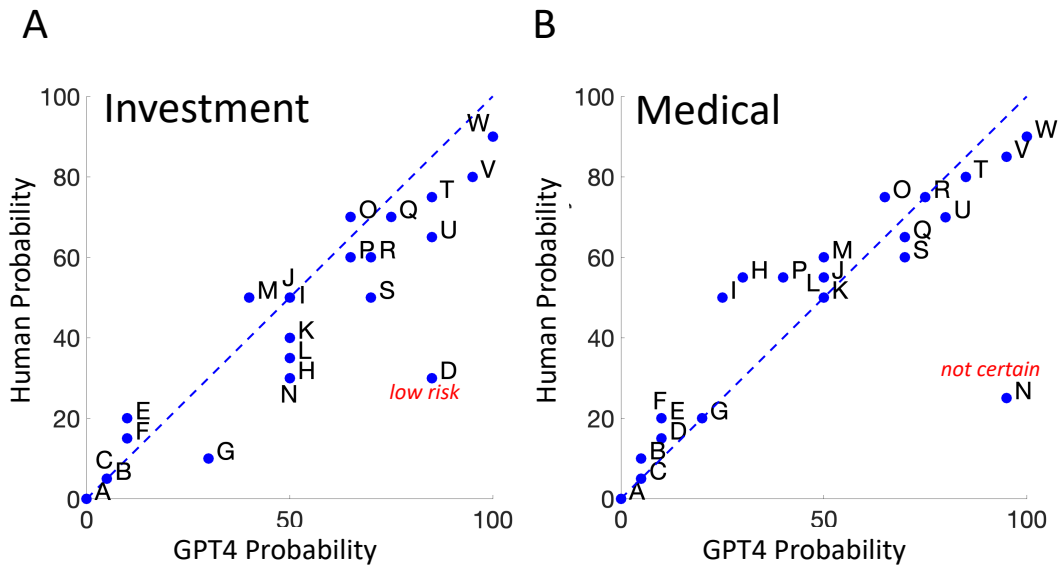

**Figure 2: Human versus GPT-4. A. Median human estimates of probability in the Investment Context are plotted versus GPT-4 estimates of probability in the same context (blue filled circles).** A letter code adjacent to the blue filled circle identifies the probabilistic phrase associated with each circle. See Table 1. One outlier (D *low risk*) is marked with its probabilistic phrase in red. See text. **B. Median human estimates of probability in the Medical Context are plotted versus GPT-4 estimates of probability.** The format is analogous to that of Figure 1A. One outlier (N *not certain*) is marked with its probabilistic phrase in red. See text.

### 1A. Human vs. GPT-4: Probability

In Figure 2 we plot the median probability ratings assigned to each of the 23 probabilistic phrases by the 25 human participants against the GPT-4 ratings of each of the 23 probabilistic phrases. Figure 2A shows results in the Investment Context while Figure 2B shows results for the Medical Context. The ratings range from 0 to 100%. If the median human participant agreed with GPT-4 in rating probability the plotted points would fall on the dashed blue identity line. The letter codes correspond to the letter codes assigned to each probabilistic phrase in Table 1. An intercept significantly different from 0 or a slope significantly different from 1 would indicate a patterned discrepancy between GPT-4 and the median human participant. We test for both possibilities.

There is an evident outlier in Figure 2A for the probability phrase "low risk" plotted in red. There is a similar outlier in Figure 2B for the probabilistic phrase "not certain". The outliers represent probabilistic phrases where GPT-4 and the median human participant assigned markedly different probabilities to the same probabilistic phrase. In the main text we report statistical analyses for this and later figures without these outliers. All results of hypothesis tests – with and without outliers – are included in a Supplement. We discuss outliers further in the section *Stability*.

***Intercept*** The Intercept estimate in Figure 2A is 1.59, not significantly[2] different from 0 [ $t(22)$ = 0.421, $p = 0.339$ ]. The Intercept estimate in Figure 2B is 11.35, significantly different from w0 [ $t(22) = 3.458$, $p = 0.0011$ ].

***Slope*** The Slope estimate in Figure 2A is 0.833, significantly different from 1 [ $t(22) = -2.577$, $p = 0.009$ ]. The Slope estimate in Figure 2B is 0.825, also significantly different from 1 [ $t(22) = -2.933$, $p = 0.0038$ ].

***Summary*** There are significant patterned differences between median human probability estimates and those of GPT-4. In both contexts median human estimates of probability are compressed by a factor of 0.8 relative to the estimates by GPT-4. In the Medical Context but not in the Investment, human estimates of probabilities are also offset vertically by roughly 10%. Human used of probability and relative frequency are typically distorted (Tversky & Kahneman, 1992; Zhang & Maloney, 2012) and the deviations we detect may be connected to probability distortion.

---

[2] We refer to tests with p-values less than 0.05 as "significant" for convenience in presenting the data. We report exact p-values for all tests in the main text and report exact p-values for all tests including those with and without outliers removed in the **Supplement**.

***Discordance.*** Both the human participants and GPT-4 are engaged in a coordination game and the second criterion of human and machine is winnings in the game. Did GPT-4 disagree with the other human human players more than they disagreed with one another?

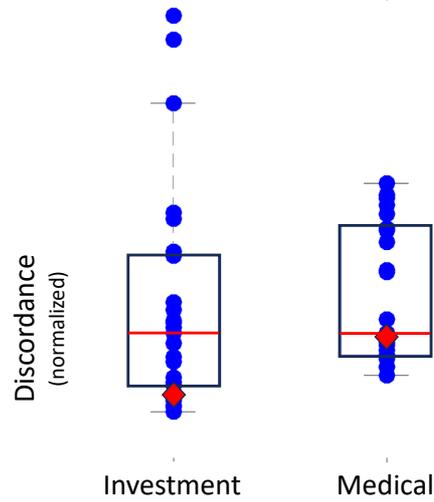

**Figure 3: Discordance.** We computed discordance, a measure of disagreement among each human observer and the remaining human observers and between GPT-4 and the human observers. See text. The left and right panels are boxplots of discordance values for the Investment Context and for the Medical Context, respectively. The top and bottom of the boxes mark the 75th and 25th percentiles for each context. The discordance for GPT-4 is marked by a solid red diamond in each context. The discordance for GPT-4 is below the median discordance (the solid red line segment) for the human participants in both contexts.

We define a measure of the disagreement between the probability or ambiguity estimates of the human participants. Let $p_i = [p_{i1},...p_{im}]$ be the vector containing the $m = 23$ probability estimates of the $i^{th}$ human participant in the order of Table 1. We define the *discordance* of the $i^{th}$ human participant to be

$$D_i = \sum_{j \sim = i} ||p_i - p_j|| \qquad (1)$$

Where $||p_i - p_j|| = \sqrt{\sum_{k=1}^{m} |p_{ik} - p_{jk}|^2}$ denotes the Euclidean distance between $p_i$ and $p_j$. The discordance of a participant is just the sum of the squared distances between the vector corresponding to the participant and each of the vectors corresponding to the remaining participants. It can be zero only if all the participants give identical estimates for all probability phrases.

Let $p_{GPT}$ denote the vector of probability estimates made by GPT-4 and define the discordance of GPT-4 to be

$$D_{GPT} = \left[\frac{m-1}{m}\right] \sum_{j=1}^{m} ||p_{GPT} - p_j|| \qquad (2)$$

There are $m-1$ summands in Eq. (1) and $m$ in Eq. (2). The multiplicative term $(m-1)/m$ in Eq. (2) corrects for the difference in the number of summands in the two equations.

Were GPT-4's judgments of probability more discrepant from those of the human participants than those of the human participants were from one another?

Figure 3A is a boxplot of all of the discordance values for the Investment Context, one blue dot per human participant. The discordance values are plotted vertically and the red horizontal line marks the median discordance in each context. The lower and upper edges of the box mark the 25$^{th}$-percentile and the 75$^{th}$-percentile, respectively. Figure 3B is the corresponding plot for the Medical Context. The red diamonds mark the discordances of GPT-4 in the two contexts. The discordance of GPT-4 is below the median of the discordances for the humans for both Contexts. ***GPT-4 agreed with the human participants as as least as well as they agreed with one another.***

*1B. Human vs. GPT-4: ambiguity*

In Figure 4 we plot the median ambiguity ratings of the 25 human participants to the GPT-4 ratings of each of the 23 probabilistic phrases. Figure 4A shows results in the Investment Context while Figure 4B shows results for the Medical Context. The letter codes once again correspond to the letter codes assigned to each probabilistic phrase in Table 1.

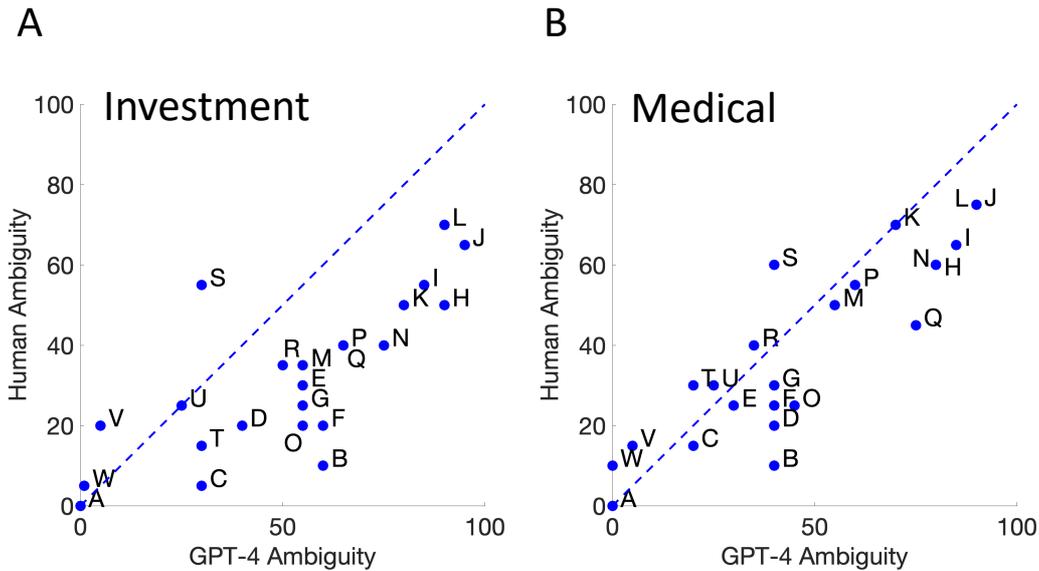

**Figure 4: Human versus GPT-4. A. Median human estimates of ambiguity in the Investment Context are plotted versus GPT-4 estimates of ambiguity in the same context as blue filled circles.** The format is analogous to that of Figure 2A. **B. Median human estimates of ambiguity in the medical context are plotted versus GPT-4 estimates of ambiguity in the same context.** The format is identical to that of Figure 2B.

*Intercept* The Intercept estimate in Figure 4A is 4.17, not significantly different from 0 [$t(23)$ = 0.729, $p = 0.237$ ]. The Intercept estimate in Figure 4B is 5.82, not significantly different from 0 [ $t(23) = 1.286, p = 0.106$ ].

*Slope* The Slope estimate in Figure 4A is 0.530, significantly different from 1 [ $t(23)$ = -4.83, $p < 0.0001$ ]. The Slope estimate in Figure 4B is 0.710, also significantly different from 1 [ $t(23) = -3.458, p = 0.001$ ].

*Summary* There are significant patterned differences between median human confidence estimates and those of GPT-4. In both contexts median human estimates are compressed, by a factor of 0.5 to 0.7 relative to the estimates by GPT-4.

We did not analyze discordance for ambiguity since there are evident large differences in estimation of ambiguity by GPT-4 and the median human participant.

*2. Comparisions across context*

We next evaluate the extent to which human judgments of probability and ambiguity are invariant across context. In Figure 5A we plot the median probability ratings of the human participants in the Investment Context versus the median probability ratings for each of the 23 probabilistic phrases of a different group of human participants in the Medical Context. A similar plot for GPT-4 is shown in Figure 5B. Figure 6A and 6B show corresponding plots for ambiguity.

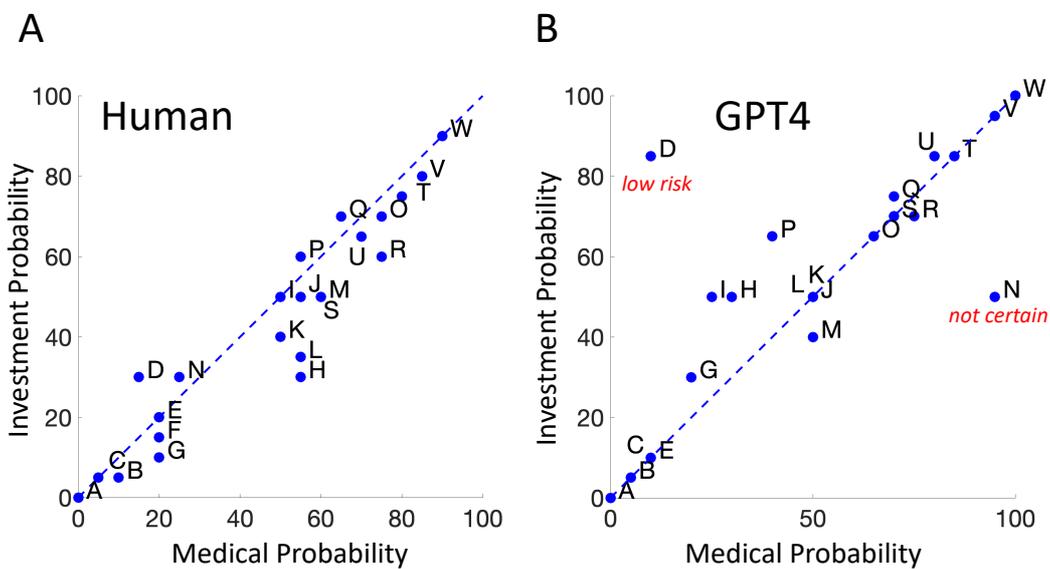

**Figure 5: The effect of context. A. Comparison of probability estimates of the median human observer across contexts.** The format is analogous to that of the corresponding panels in Figure 2A**.** All data fall roughly along the identity line. Human participants select lower probabilities for the same probabilistic phrase in the Medical Context. **B. Comparison of GPT-4 probability estimates across contexts.** The format is analogous to that of Figure 2B. All data fall roughly along the identity line. The same two outliers appear in Figure 4B as in Figure 2A and Figure 2B.

***Intercept*** The Intercept estimate in Figure 5A is -0.147, not significantly different from 0 [t(23) = -0.041, p = 0.484 ]. The Intercept estimate in Figure 5B is -2.41, not significantly different from 0 [ t(23) = -0.608, p = 0.275 ].

***Slope*** The Slope estimate in Figure 5A is 0.907, is not significantly different from 1 [ t(23) = -1.41, p = .0865 ]. The Slope estimate in Figure 5B is 0.977, not significantly different from 1 [ t(21)  = -0.341, p = 0.368 ].

***Summary*** There are no significant patterned differences between median human probability estimates in the Investment Context and the Medical Context. There are no significant patterned differences between estimates by GPT-4 in the Investment Context and the Medical Context those of GPT-4.

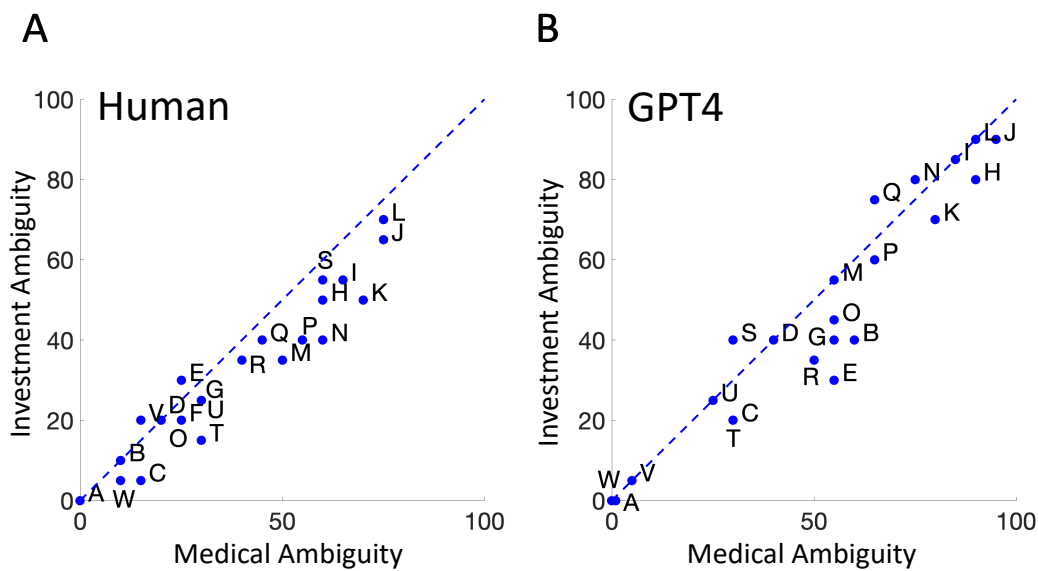

**Figure 6:The effect of context. A. The median of human estimates of ambiguity in the Investment Context are plotted versus the median of human estimates of ambiguity in the Medical Context.** The format is analogous to that of Figure 5A. Human estimates are significantly higher in the medical context. See text. **B. The median of GPT-4 estimates of ambiguity in the Investment Context are plotted versus the median of GPT-4 estimates of ambiguity in the Medical Context.** GPT-4 estimates are significantly higher in the medical context. See text.

***Intercept*** The Intercept estimate in Figure 6A is 0.048, not significantly different from 0 [t(23) = 0.021, p = 0.492 ]. The Intercept estimate in Figure 6B is 8.42, significantly different from 0 [ t(23) = 2.214, p = 0.019 ].

***Slope*** The Slope estimate in Figure 6A is 0.819, significantly different from 1 [ t(23) = -3.46 p= .0010 ]. The Slope estimate in Figure 6B is 0.941, not significantly different from 1 [ t(23) = -0.341, p = 0.207 ].

***Summary*** There are significant patterned differences between ambiguity estimates by GPT-4 in the Investment Context and the Medical Context those of GPT-4.

*3. Stability*

The two outliers in GPT-4's performance raise issues concerning the stability of GPT-4. We chose to examine the outlier in Figure 2B (the probabilistic phrase *not certain*) to determine whether it reliably recurs (representing a large but reliable discrepancy between human and GPT-4 estimates) or whether it is evidence of instability. If it reliably reoccurs then it is effectively a difference of opinion between human and machine as to the meaning of a particular probabilistic phrase. If not, it would suggest that GPT-4 is unstable.

GPT-4 included explanations for its responses. We tabulate these explanations for the outlier in Figure 2A (***low risk***) in the Investment Context and the outlier in Figure 2B (***not certain)*** in the Medical Context. The reader may agree with GPT-4 or not, but GPT-4's response acknowledges that the probabilistic phrases can be interpreted in more than one way and perhaps human and machine are simply in disagreement.

Will the outlier recur if we rerun the trial? We redid the GPT-4 estimates in the Medical Context four times[3], plotting the estimates as four blue contours in Figure 7. The original estimates are plotted in red with a red solid circle marking the outlier. In brief, we did not reproduce the anomalous outlier we initially encountered nor did other outliers emerge for any of the other probabilistic phrases. The four new estimates of probability are in good agreement

---

[3] The GPT-4 interface that we have access to limits the number of runs that we can carry out in a fixed period of time, precluding analyses that require large numbers of repetitions of trials.

with those of the human participants and with one another but not with the original GPT-4 estimate.

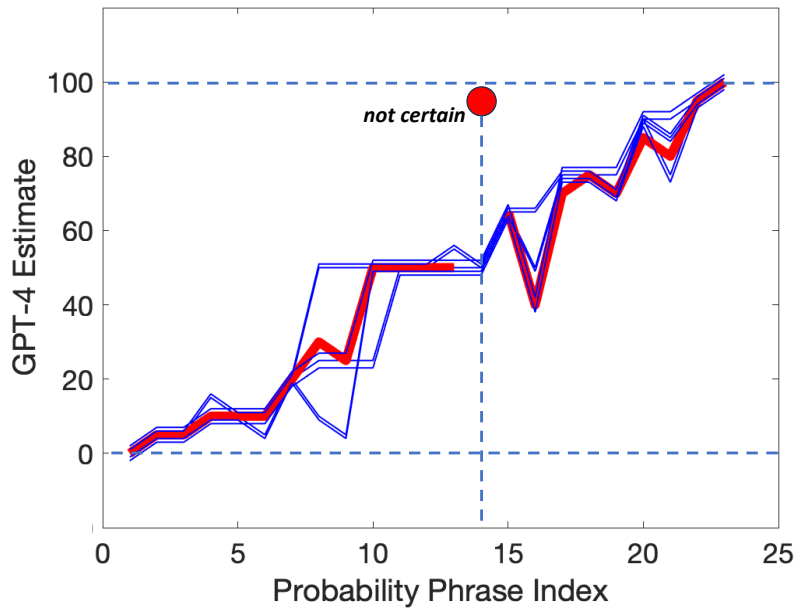

**Figure 7: Stability of GPT-4 estimates in the Medical Condition.** Figure 2B (GPT-4 estimates of probability in the Medical Context) contains an evident outlier *not certain* whose estimate in the Investment Condition is 50%, in the Medical Context, 95%. We plot the GPT-4 estimates from the Medical Context in red with the outlier marked by a red disk. We repeat these GPT-4 estimates four times and plot them as well, in blue. The contours are slightly offset vertically so that they are visible. The outlier does not reoccur and the four repetitions are in good agreement with each other and the original estimates except for the outlier. See text.

# DISCUSSION

In coordination games we use words to share information in order to coordinate actions (Benz, Ebert, Jäger & Van Rooij 2011; Benz, Jäger, & Van Rooij, 2014). The specific coordination game we consider here concerns correct use of probabilistic phrases signaling probability and ambiguity. There were two versions of the game, one with probabilistic phrases used to give investment advice, one with these same phrases used to give medical advice. Half the human participants ran in the Investment Context, half in the Medical.

**Conclusions: Estimates of Probability**

- As measured by discordance, **GPT-4 agreed with human participants as least as well as the median human participant agreed with the other participants**. Based on overall performance we cannot distinguish GPT-4 and human (Figure 3).
- Examined in detail (residuals analysis, Stigler, 2016), we found significant patterned discrepancies between GPT-4's estimates of probability and those of human participants that could be captured by fits to a linear model (Figure 2). In both contexts, human estimates of probabilities tended to be compressed relative to those of GPT-4.
- Use of probabilistic phrases by both GPT-4 and humans signaling probability transferred well across contexts for the two contexts we considered. (Figures 5). A doctor's use of *likely* conveys the same information about probability as that of an investment consultant.

**Conclusions: Estimates of Ambiguity**

- Human estimates of ambiguity were compressed relative to those of GPT-4 by roughly a factor of 2. However, unlike probability, there is no standard scale of ambiguity. We can only conclude that GPT-4 did not anticipate human use of the ambiguity scale, a failure to coordinate.

There is some indication that GPT-4 is unstable, producing occasional outliers. Further research is needed to evaluate this apparent instability.

| Context | Investment | Medical |
|---|---|---|
| **"low risk"** | Estimate: **85%** | Estimate: **10%** |
| | **Justification**: When you describe something as "low risk" in the context of investments increasing in value, you're conveying that there's a high likelihood the investments will perform positively or, at the very least, not decline significantly. Given the options provided, and interpreting "low risk" as a high probability of the investments increasing in value, I would estimate the communicated probability to be: 85%. This reflects that there's a high likelihood of positive performance, but as with all investments, there's never a guarantee. Always remember that while terms like "low risk" are commonly understood in finance, their interpretation can still vary somewhat based on individual perception and context." | **Justification**: 'If I use the phrase "low risk" to describe the probability of a headache occurring as a side effect, I'm suggesting that it's not a common outcome, but there's still some possibility. Given the list of percentages you provided, "10%" seems to be a good representation of what "low risk" might imply in this context." |
| **"not certain"** | Estimate: **50%** | **95%** |
| | The phrase "not certain" is somewhat ambiguous. While it indicates that there isn't a guarantee, it doesn't provide a clear indication of how likely or unlikely the event is. Given its vagueness, but still leaning toward the idea that there's less confidence than "certain," a neutral probability might be represented. Therefore, I would estimate the communicated probability for "not certain" to be: 50%. However, it's essential to understand that this phrase leaves a lot to interpretation, and different people might infer different probabilities from it. | The phrase "not certain" is ambiguous and indicates that there's uncertainty, but it doesn't necessarily provide a clear indication of how probable or improbable the event is. It could be slightly probable or highly probable, but just not 100% certain. Given the options provided and the context, "95%" might be the best choice to convey that there's a high likelihood, but it's not certain. However, it's important to note that the phrase "not certain" doesn't inherently align with any specific percentage; the chosen value is more of an educated guess based on the given options. |

**Table 2: GPT-4's Justifications of Estimated Probabilities.** GPT-4's probability estimates for the two different contexts were discrepant for probabilistic phrases "low risk" and "not certain". GPT-4's output includes an explanation of each choice which we include here.

We focused on one particular coordination game and compared human and machine. Similar games could be based on color terms or sets of dimensional adjectives (Faller, 1999), for example, the dimensional adjectives describing size: *small*, *big*, *large, etc.* (Bierwisch, 1967; Maloney and Gelman, 1987). Gurnee & Tegmark (2023) look at representation of space and time.

But human use of probability phrases is a particularly rich source of possible coordination games that we could use to compare human and machine. We can challenge GPT-4 to play each game we develop, comparing human and machine as we did here.

When, for example, do humans use probability phrases and when do they use numerical probability? Dhami & Mandel (2022) in their review article argue that the choice between the use of numerical or verbal probabilities by senders is influenced by several factors. For example, Juanchich and Sirota (2020) find that, in the medical context, senders prefer to use numerical values when uncertainty is about very consequential events, as, for example, the serious side effects of a drug. Would GPT-4 have similar preferences?

Wallsten, Budescu, Zwick & Kemp (1993) found that most people preferred to receive information about the probability of a chance event in numerical form but preferred to transmit this information as a probabilistic phrase. Erev & Cohen (1990) referred to this pattern of preference as the Communication Mode Preference Paradox. Whatever justification we offer for transmitting a probabilistic phrase instead of a numerical probability would seem to apply to receiving it in the same form, an apparent paradox. Would GPT-4 exhibit the same paradox?

Senders' use of verbal probabilities has several effects other than conveying an estimate of uncertainty (Dhami & Mandel, 2022). Collins and Mandel (2019) suggest that expressing probabilities verbally conveys implicit recommendations of actions. Indeed, Honda et al. (2017) found that the use of verbal probabilities, for example using positive rather than negative terms, can introduce bias in the decison making process. Verbal probabilities can also be used as receiver's and sender's face-saving strategy (Bonnefon and Villejoubert, 2006; Juanchich. et al., 2012; Jenkins and Harris, 2020). Does GPT-4 exhibit similar biases?

Comparing GPT-4 to human in these coordination games provides a systematic way to assess what GPT-4 (or any other LLM) can and cannot do, where its performance matches, exceeds or falls short of human. There is more to language than a series of coordination games but such games provide a scaffolding allowing us to describe what GPT-4 does in a principled way.

One puzzle we are left with is the compressions of probability and ambiguity we found. Despite these failures, GPT-4's overall performance measured by discordance is better than that of the typical human participant. Yet a simple scaling of output would "fix" the compression problem and presumably improve its performance. The methods used to develop GPT-4 did not result in an LLN that included appropriate scalings. Why?

# METHODS

***GPT-4.*** Typical inputs for GPT-4 are shown in Table 3 for both contexts and for estimates of probability and ratings of ambiguity. Input consisted of a single context phrase followed by a request for a rating of either the probability or ambiguity of a specified probabilistic phrase. The phrases were presented in randomized order. GPT-4 was constrained to respond with a single percentage (probability) or number (ambiguity) and a justification. The values permitted were multiples of 5% from 0% to 100% (0%, 5%, …, 100%) and multiples of 5 from 0 to 100 for ambiguity ratings. We analyzed only the first run of GPT-4 in Figures 1-6.

***Human participants.*** Fifty participants drawn from the New York University SONA Subject Pool agreed to participate in the experiment. All completed the experiment. We report the demographics of the pool. Almost all of the potential participants in the pool were between the ages of 18-35 (93.34%), only a small portion of the participants were over 35 (6.38%) or under 18 (0.27%). 30.9% of the potential participants were male at birth and 69.0% were female at birth. The research protocol and methods were approved by the Institutional Review Board for that Faculty of Arts and Science at New York University (IRB-FY2023-7544).

***Procedure.*** Each participant was taken to a laboratory room and seated in front of a computer screen. Each participant was assigned at random to one of two contexts, investment or medical, 25 participants in each. Stimuli were presented and responses recorded using Google forms.

| A Investment Probability | B Medical Probability |
|---|---|
| Imagine you are a financial consultant trying to convey to a client the probability that their investments will increase in value over the next year. If you use the phrase '***almost certain'*** what probability are you trying to communicate?<br><br>*[Scale from 0% to 100% in steps of 5%.]* | Imagine you are a doctor trying to convey to a patient the probability of a headache occurring as a side effect from a drug you had prescribed. If you use the phrase '***almost certain'*** what probability are you trying to communicate?<br><br>*[Scale from 0% to 100% in steps of 5%.]* |
| **C Investment Ambiguity** | **D Medical Ambiguity** |
| Imagine you are a fin<br>ancial consultant trying to convey to a client the probability that their investments will increase in value over the next year. You use the phrase '***almost certain'.*** How ambiguous is that term?<br><br>*[Scale from 0 to 100 in steps of 5.]* | Imagine you are a doctor trying to convey to a patient the probability of a headache occurring as a side effect from a drug you had prescribed. You use the phrase '***almost certain'.*** How ambiguous is that term?<br><br>*[Scale from 0 to 100 in steps of 5.]* |

**Table 3: Sample input for GPT-4 for the probabilistic phrase "almost certain" as presented in four conditions.** For each condition we constructed 23 stimuli for the 23 probabilistic phrases in Table 1. Participants rated the probability (0%, 5%, 10%, …, 100%) and the ambiguity (0, 5, 10, …, 100) of each phrase (0 = least ambiguous, 100 = most ambiguous). Ambiguity ratings were collected after probability ratings. The scale values were enumerated explicitly, 0, 5, … 100. Human participants completed A and then C or B and then D but not both. GPT-4 ran A and then C and was then restarted (erasing knowledge of trials completed) before completing B and D. See text.

Participants assigned to the medical context answered questions about probability phrases and ambiguity of probability phrases related to medical advice (See Table 3). The questions were prefaced by a single sentence signaling context as were those posed to GPT. Participant assigned to the investment context answered questions related to investment advice (See Table 3).

Questionaires in both contexts consisted of same two sections: probability and ambiguity. Section 1 asked subjects to rate the probability of 23 words and phrases on a percentage scale from 0 to 100% (with 5% increments). Section 2 asked participants to rate the ambiguity of the same 23 phrases using a scale ranging from 0 to 100 in steps of 5. Questions within each section were randomly re-ordered for each participant.

*Hypothesis tests*: In each panel of Figures 2, 4, 5, 6 we plotted corresponding data (e.g. estimates from human participants and corresponding estimates from GPT-4) as a scatterplot, allowing the reader to assess the relations between variables visually (Mosteller & Tukey, 1977). We fit the data in each scatterplot by a univariate linear model. The null hypothesis for each test was that the points fell on the identity line with slope 1 and intercept 0 with added iid Gaussian error. We used hypothesis tests to detect deviations of slope from 1 or intercept from 0) allowing us detect any trends in the data captured by the estimates of intercept and slope, a form of residuals analysis (Stigler, 2016). We discuss any significant trend captured by the estimated slope and intercept values (compression or offset).

All tests were two-tailed with size 0.05. We report the t-statistic, the degrees of freedom and the exact p-value for each test and refer to outcomes with p-value less than 0.05 as "significant" in discussing the data. We classified two data points as outliers, and label them where they appear in the scatterplots by their probabilistic phrases in red. We report results with outliers excluded in the main text and report the analyses of all tests with outliers excluded and with outliers included in a **Supplement**.